\documentclass[12pt]{article}
\usepackage{a4}
\usepackage{graphicx}

\def\lsim{\;
\raise0.3ex\hbox{$<$\kern-0.75em\raise-1.1ex\hbox{$\sim$}}\;
}

\def\gsim{\;
\raise0.3ex\hbox{$>$\kern-0.75em\raise-1.1ex\hbox{$\sim$}}\;
}

\begin{document}

\renewcommand{\thefootnote}{\fnsymbol{footnote}}

\title{{\bf \large Photoproduction of quasi-bound \boldmath $\omega$\unboldmath
\, mesons in nuclei}\thanks{Work supported in part by DFG.}
}

\author{E. Marco\footnotemark[7] ~ and W. Weise\\
\\
Physik-Department,\\
Technische Universit\"at M\"unchen,\\
D-85747 Garching, Germany}

\maketitle

\footnotetext[7]{Fellow of the A.v.Humboldt Foundation}

\vspace{-10cm}
\begin{flushright}
{\bf TUM/T39-00-23}
\end{flushright}
\vspace{8cm}

\begin{abstract}
We propose the $(\gamma,p)$ reaction as a means of
producing possible quasi-bound states of $\omega$ mesons in nuclei.
We use an effective Lagrangian, based on chiral $SU(3)$
symmetry and vector meson dominance, in order to construct
the $\omega$-nuclear potential. The
contribution of bound $\omega$ states is observable in the missing energy
spectra of protons emitted in forward direction for several nuclei.
\end{abstract}



\newpage

   The behaviour of vector mesons in the nuclear medium is
one of the topics which attracts much attention in current
nuclear physics. At high temperatures (and possibly at extremely high
densities), the chiral symmetry of QCD is expected
to be restored and vector and axial vector mesons become degenerate.
At moderate densities characteristic of nuclei, one expects
to see a downward shift of the spectral distributions of
vector mesons. QCD sum rules together with current algebra considerations
\cite{FESR,Peris} suggest that the first moment of vector
meson mass spectra should be linked to the ``chiral gap'',
$4 \pi f_{\pi} \sim 1$ GeV, and one expects the pion decay constant
$f_{\pi}$ to decrease with increasing baryon density, roughly like
the square root of the chiral (quark) condensate.

Several studies, mostly concerned with the $\rho$ meson spectrum
and its possible implications for dilepton spectra produced in
heavy-ion collisions, hint at a decrease of vector meson masses in
the nuclear medium. The study of QCD sum rules in the medium \cite{Hatsuda}
and the Brown-Rho scaling hypothesis \cite{BrownRho} suggest a
dropping of the vector meson masses by approximately 15\% at
normal nuclear matter density. Dynamical studies of the behaviour
of vector mesons in the medium predict a considerable increase of
the $\rho$ width, but no significant decrease in its
mass \cite{Chanfray,RapWam,KlinglKaiser}. For the $\omega$ meson
\cite{Klingl-ome}, such models predict a decrease of
its mass accompanied by a moderate increase of its width, so that
the $\omega$ appears to be a better candidate for in-medium
studies than the $\rho$.

The $(d,^3\!\mbox{He})$ reaction with recoilless kinematics
has been proposed in order to search for possible $\eta$
and $\omega$ meson bound states in nuclei \cite{Klingl-ome,Hayano,Saito}.
This reaction allows the study of the in-medium behaviour of mesons
under well controlled conditions, complementary to heavy-ion collisions.

In this paper we propose the use of the $(\gamma,p)$ reaction in
nuclei to explore the behaviour of the $\omega$ meson in the nucleus. 
The difference between the $(d,^3\!\mbox{He})$ and $(\gamma,p)$
reactions is that in the latter case, distortion effects
are restricted only to the proton in the final state.
The optimal energies needed to produce the $\omega$ at
rest, around 2.75 GeV in the photoproduction case, are available
at several experimental facilities such as ELSA in Bonn and
Spring8 in Japan. 

   In our calculation we have used the model developed in 
\cite{KlinglKaiser,KlinglZPA,KaiserWaas}
to generate the (complex) potential experienced by
the $\omega$ in the nucleus. It is derived using an effective
Lagrangian which combines chiral $SU(3)$ and vector meson dominance.
One of the important features of the model
is that the self-energy of the $\omega$ in the medium is strongly energy
dependent, in contrast to the calculations of \cite{Hayano},
where a static potential is used. At
nuclear matter density the $\omega$ mass is reduced by approximately
15 \% and its width is increased up to around 40 MeV. Unlike the
$\rho$ meson with its prohibitively large
in-medium width, the $\omega$ has chances of being
observed as a quasiparticle state in the nucleus.

   Our choice of the kinematic conditions is such that 
the $\omega$ is produced practically at rest in the nucleus.
Its longitudinal and transverse self-energies are therefore nearly the same
\cite{KlinglKaiser}, and it is justified to work with a single
scalar self-energy $\Pi(E,\vec{r})$. In the
local density approximation, the potential that the $\omega$ experiences
in the nuclear medium can be expressed as

\begin{eqnarray}
\Pi(E,\vec{r}) &\equiv& 2 E U(E,\vec{r})\nonumber\\
&=&-\mbox{Re} T_{\omega N} (E) \rho(\vec{r}) - i \left[ 
E \Gamma_{\omega}^{(0)} (E) + \mbox{Im} T_{\omega N} (E) \rho(\vec{r})
\right ] \, , \label{eq:potential}
\end{eqnarray}
where, $\Gamma_{\omega}^{(0)} (E)$ is the free $\omega$ decay width, and
$T_{\omega N} (E)$ is the energy-dependent $\omega$-nucleon amplitude
in free space, which we take from \cite{KlinglKaiser}.

   To evaluate the $\omega$ bound states one must solve
the wave equation with the potential (\ref{eq:potential}),

\begin{equation}
\left[ E^2 + \nabla^2 - m_{\omega}^2 - \Pi(E,\vec{r})\right]
\phi(\vec{r}) = 0 \, ,
\label{eq:waveeq}
\end{equation}
self-consistently
to obtain the (complex) quasi-bound state energies $E_{\lambda}$.
We use a two parameter Fermi distribution to describe the nuclear
density distributions. The quasi-bound $\omega$ states
found for different nuclei are tabulated in
Table 1, where we have introduced

\begin{table}
\begin{center}
\begin{tabular}{ccccc} \hline
	&	&	& $(\varepsilon_{nl},\Gamma_{nl})$ [MeV]&\\ \cline{3-5}
nucleus	& n	& $l=0$	& $l=1$	& $l=2$\\ \hline
$^{\,6}_{\omega}\mbox{He}$ & (1) & $(-49,36)$ & $(-18,33)$ & $-$\\
\\
$^{11}_{\omega}\mbox{B}$ & (1) & $(-66,41)$ & $(-40,39)$ & $(-13,37)$\\
		& (2) & $(-14,34)$ & $-$ & $-$\\
\\
$^{39}_{\omega}\mbox{K}$ & (1) & $(-88,44)$ & $(-73,45)$ & $(-57,45)$\\
		 & (2) & $(-54,45)$ & $(-36,44)$ &$(-17,44)$ \\
		& (3) & $(-16,41)$ & $-$ &$-$ \\ \hline
\end{tabular}
\end{center}
\caption{Complex energy eigenvalues (see Eqs.\
(\ref{eq:energy},\ref{eq:gamma}) of an $\omega$ meson bound to
several nuclei.}
\end{table}

\begin{equation}
\varepsilon_{\lambda} = \mbox{Re}E_{\lambda} - m_{\omega}\, ,
\label{eq:energy}
\end{equation}
and the total decay widths

\begin{equation}
\Gamma_{\lambda} = -2 \, \mbox{Im} E_{\lambda}\, .
\label{eq:gamma}
\end{equation}
One observes that the $\omega$ is bound even in light nuclei such as
$^6$He. The widths are around $35\sim 45$ MeV. Their origin is
primarily the reaction $\omega N \rightarrow \pi N$ in the nucleus,
as calculated in ref.\ \cite{KlinglKaiser}. Although these widths
prohibit identifying individual peaks for each state, one should
nevertheless be able to observe strength at energies below the
threshold for quasifree $\omega$ production.
	
The reaction that we propose in order to detect bound $\omega$
states is $(\gamma, p)$ on nuclei. At an incoming photon energy of
around 2.75 GeV, and with the proton emitted in forward direction,
the $\omega$ is produced nearly at rest. One can scan the
contributions of the bound $\omega$ states by observing the missing
energy spectrum, i.e. by measuring the energy of the outgoing proton in
the forward direction and plotting the differential cross section
as a function of $E_{\omega}  - m_{\omega} + |B_p| = E_{\gamma} 
+m_p - E_{p} - m_{\omega}$, where $E_{p}=m_p + T_p$ is the detected
energy of the outgoing proton, and $B_p$ is the binding energy
of the bound initial proton.

\begin{figure}[t]
\centerline{
\includegraphics[width=0.8\textwidth]{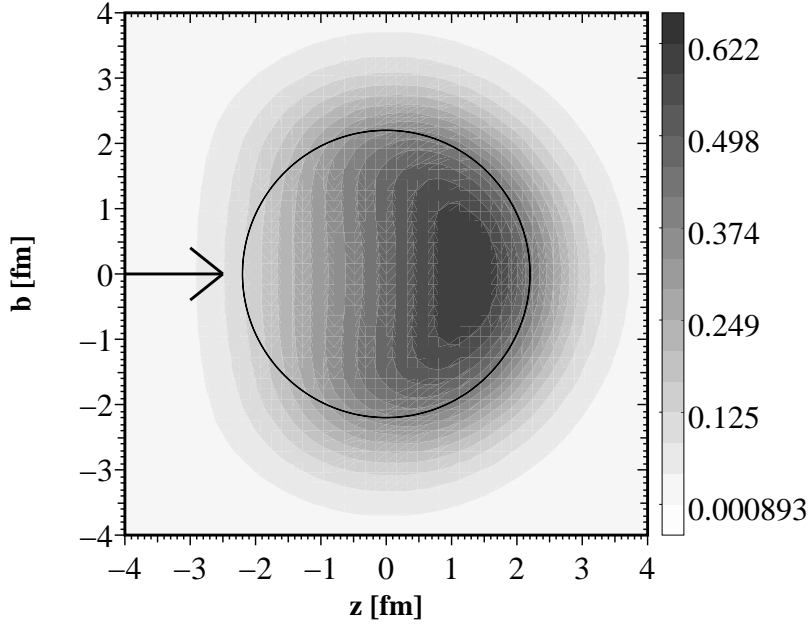}
}
\caption{Values of $D(z,b)\rho(r)/\rho(0)$ for the $(\gamma, p)$ reaction
on $^{12}$C. The reaction takes place predominantly at the
darker regions. The circle indicates the r.m.s. radius of $^{12}$C}
\label{fig:distortion}
\end{figure}

In order to evaluate the cross section we use the distorted
wave impulse approximation method (DWIA) \cite{DWIA1,DWIA2,DWIA3}.
The nuclear $\omega$ meson photoproduction cross section,
with the proton emitted at zero angle, is expressed as

\begin{equation}
\left(\frac{d^2\sigma_{\gamma + A \rightarrow p + \omega(A-1)}}
{dE d\Omega}\right)^{\mbox{\scriptsize lab}}_{\theta_p=0}
=
\left(\frac{d\sigma_{\gamma + p \rightarrow p + \omega}}
{d\Omega}\right)^{\mbox{\scriptsize lab}}_{\theta_p=0}
S(E) \, .
\end{equation}
For the free cross section, $d\sigma_{\gamma + p \rightarrow
p + \omega}/d\Omega$, we take a value of 0.3 $\mu$b/sr from 
ref.~\cite{SAPHIR}.

$S(E)$ is the response function, which takes into account the removal
of the proton from the nucleus, the binding of the $\omega$ meson in the
nucleus and the distortion of the outgoing proton wave. The
following expression holds for $S(E)$:

\begin{eqnarray}
S(E)&=&\sum_{j_p,l_p}\sum_{l,L}N_p\frac{2l+1}{4\pi} (l_p0l0|L0)^2 \nonumber \\
&\times&\mbox{Im} \int^{\infty}_0 dr' \,r'^2 w_L^*(r') \psi^*_{j_p l_p} (r')
\int^{\infty}_0 dr \, r^2 w_L(r) \psi_{j_p l_p} (r)
g_l (E-B_p,r',r)\, . \nonumber
\end{eqnarray}
Here $\psi_{j_p l_p} (r)$ is the radial wave function of the initial
bound proton,

\begin{equation}
g_l (E-B_p,r',r) = 2 i E u_l(k,r_<) v^*_l(k, r_>)
\end{equation}
is the radial Green function of the $\omega$ meson for a given
angular momentum $l$, expressed in terms of the regular and outgoing
solutions of the wave equation (\ref{eq:waveeq}), and

\begin{equation}
w_L(r) = \int^1_{-1} d \, \cos (\Theta) e^{i (p_{\gamma}-p_p) r \cos \Theta}
D(z(\Theta), b(\Theta)) P_L (\cos \Theta)\, ,
\end{equation}
with Legendre polynomials $P_L$. The distortion
factor $D(\vec{r})$ is evaluated using the eikonal approximation
for the wave function of the outgoing proton:

\begin{equation}
\psi^{\dag}_f (\vec{p}_p,\vec{r}) = e^{-ip_p z} D (\vec{r})\, .
\end{equation}
This approximation is justified, given that the proton has
a kinetic energy of $1\sim 2$ GeV in our cases of interest. 
In terms of $z = r \cos \Theta$ and the impact parameter $b$, the distortion
factor has the form

\begin{equation}
D(\vec{r}) = \exp \left[ - \frac{\sigma_{p N}}{2} \int_z^{\infty}
dz' \, \rho(z', b)
\right].
\end{equation}
For the proton-nucleon cross section we have taken a value
$\sigma_{pN} = 40$ mb. Fig.\ \ref{fig:distortion} shows
$D(\vec{r}) \rho(\vec{r})/\rho(0)$ in order to illustrate the
``active'' zone of the process. One
observes that the reaction takes place predominantly in the rear
hemisphere of the nucleus, since the protons are distorted on their
way out. In the case of the $(d,^3\!\mbox{He})$ reaction,
only the edges of the nuclear surface are actively involved
\cite{Klingl-ome}, because the incoming deuteron
and the outgoing $^3$He are both strongly absorbed. As a
consequence, the reduction of the  $(d,^3\!\mbox{He})$ cross
section for $\omega$ production is about 10 times stronger than in the
photoproduction case.

\begin{figure}[t]
\centerline{
\includegraphics[width=0.6\textwidth,angle=-90]{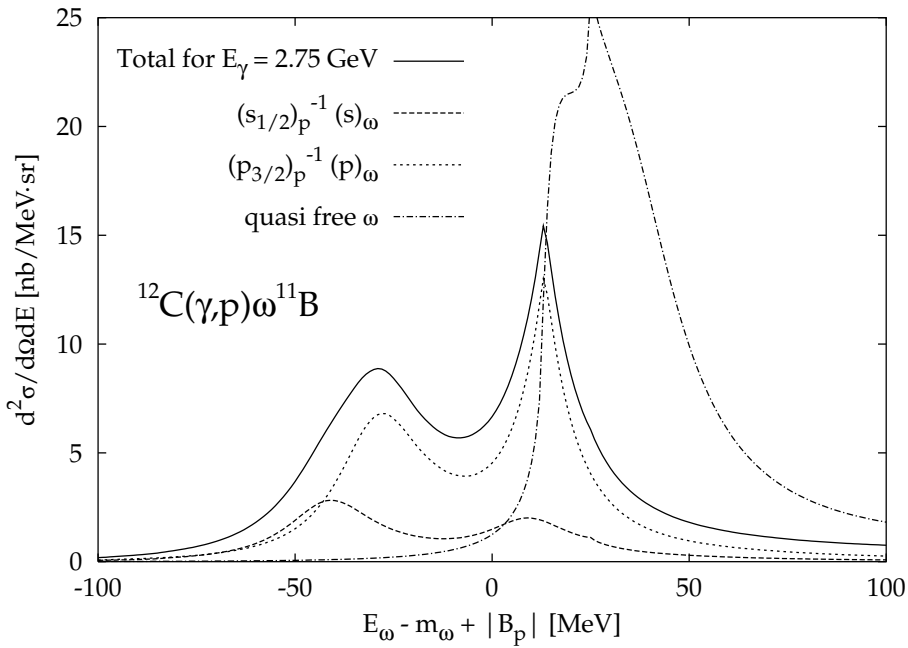}
}
\caption{Missing energy spectra for the $^{12}\mbox{C}(\gamma, 
p)\omega^{11}\mbox{B}$ reaction at $E_{\gamma}=2.75$ GeV.
Dotted lines represent the contributions from
two particular combinations of bound $\omega$ and proton-hole
states.}
\label{fig:C2.75}
\end{figure}

In Fig.\ \ref{fig:C2.75} we show the calculated proton missing energy
spectrum for the $^{12}\mbox{C}(\gamma, p)\omega^{11}\mbox{B}$ reaction
with an incoming photon energy of 2.75 GeV. This is the energy at which
a free $\omega$ would be produced at rest. Pronounced 
structures coming from different bound $\omega$ states
can be seen below the threshold for quasi-free
$\omega$ production. The figure also shows separately
the contributions of two of the more prominent combinations of $\omega$ and
proton states. The prominent structure seen at
$E_{\omega}  - m_{\omega} + |B_p| \simeq 13$ MeV is characteristic
of ${\omega}^{11}\mbox{B}$ with the $\omega$ and the initially bound
proton in $p$-orbitals, and reflects a threshold effect.

\begin{figure}[!h]
\centerline{
\includegraphics[width=0.6\textwidth,angle=-90]{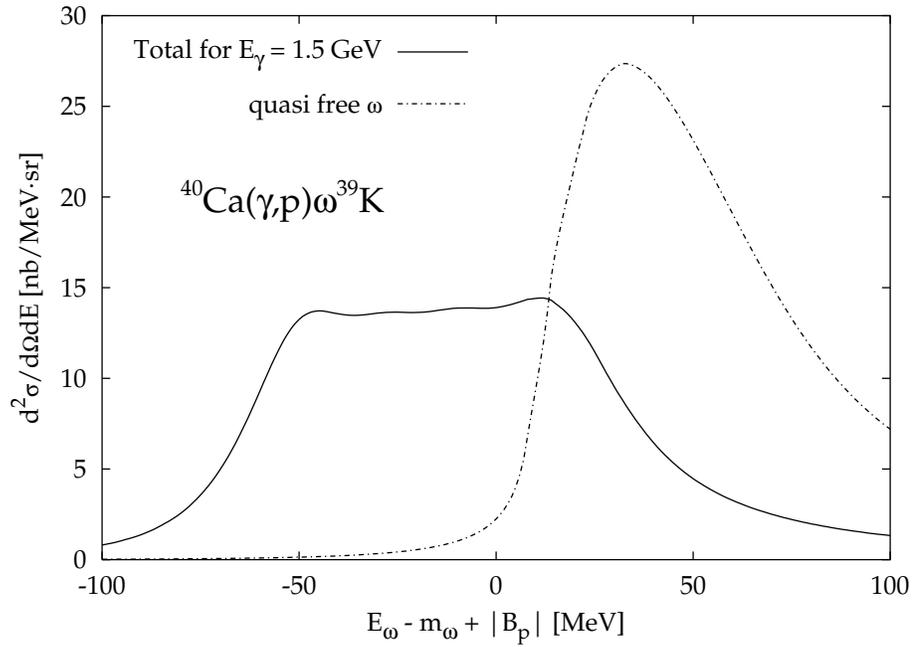}
}
\caption{Missing energy spectra for the 
$^{40}\mbox{Ca}(\gamma, p)\omega^{39}\mbox{K}$
reaction at $E_{\gamma}=1.5$ GeV.}
\label{fig:Ca1.5}
\end{figure}

\begin{figure}[!h]
\centerline{
\includegraphics[width=0.6\textwidth,angle=-90]{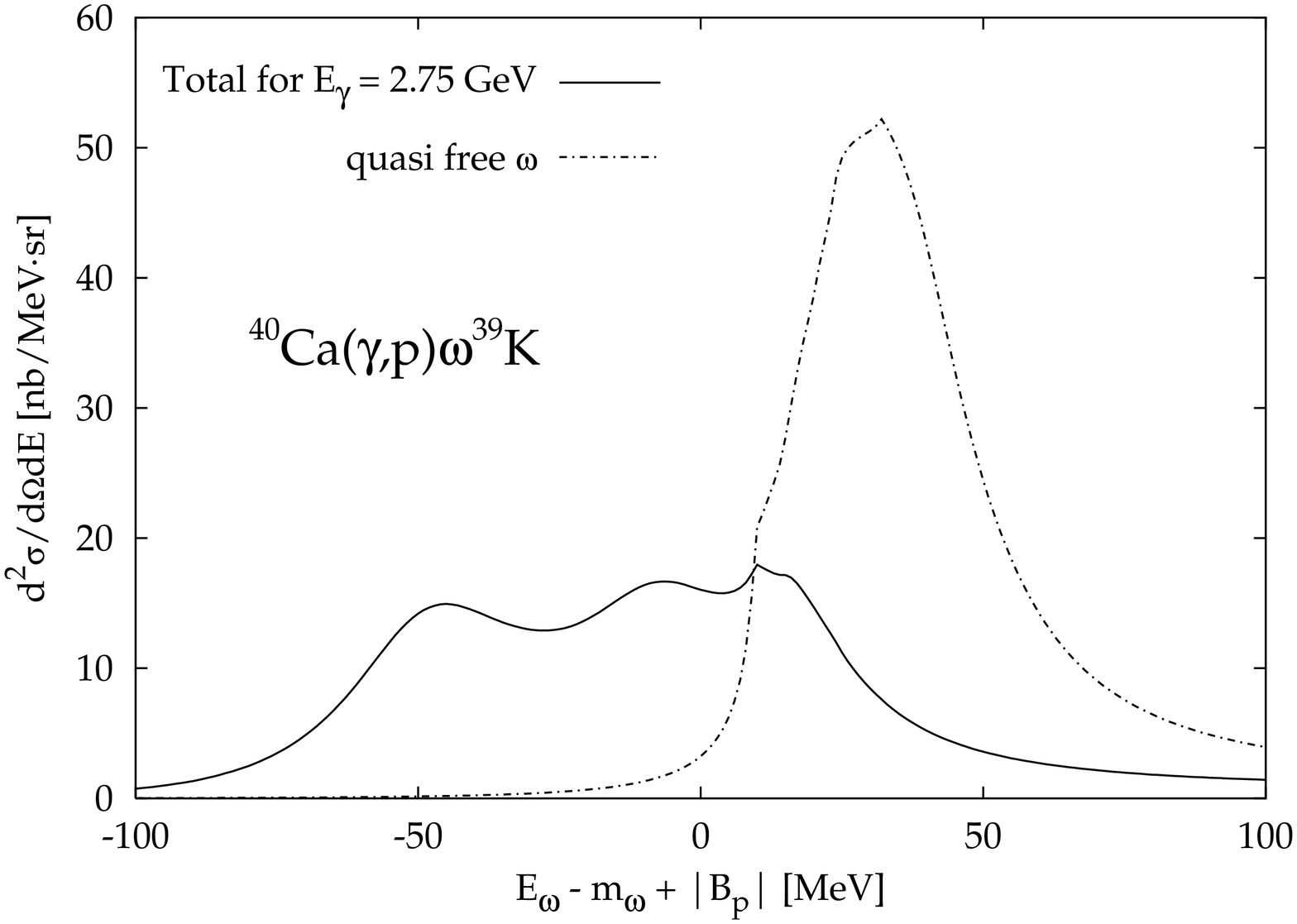}
}
\caption{Missing energy spectra for the 
$^{40}\mbox{Ca}(\gamma, p)\omega^{39}\mbox{K}$
reaction at $E_{\gamma}=2.75$ GeV.}
\label{fig:Ca2.75}
\end{figure}

The results for the $(\gamma, p)$ reaction on $^{40}$Ca at
$E_{\gamma}=1.5$ GeV and 2.75 GeV are shown in Figs.\ 
\ref{fig:Ca1.5} and \ref{fig:Ca2.75} respectively. In both
cases there are important contributions coming from the bound
$\omega$ mesons. For $E_{\gamma}=1.5$ GeV, a free $\omega$ meson
would have a momentum of around 130 MeV/c, comparable to that in the
suggested $(d,^3\!\mbox{He})$ experiments \cite{Klingl-ome,Hayano},
at $T_d = 4$ GeV. The cross section at $E_{\gamma}=2.75$ GeV is slightly
higher than at $E_{\gamma}=1.5$ GeV.

If only the missing energy of the recoiling proton is detected,
the spectrum of produced quasibound $\omega$ mesons is expected
to sit on a background, the cross section of which may be approximately
five times as large as the $\omega$ meson signal itself. This background
should be flat, resulting primarily from the $(\gamma, p)$ reactions
leading to $\rho$ meson and continuum $\pi \pi$ production,
with the $\rho$ meson width strongly increased in the presence
of the nucleus. Ideally, the background would be reduced by detecting
a characteristic decay mode of the $\omega$ meson together with the
forward proton.

\hspace{1cm}

We thank Albrecht Gillitzer, Satoru Hirenzaki,
Paul Kienle, Eberhard Klempt, Berthold Schoch and Hiroshi Toki for helpful
discussions.

\newpage  

\end{document}